\documentclass[useAMS,usenatbib]{mn2e}
\usepackage{epsfig}
\input psfig.sty

\usepackage{graphicx}% Include figure files
\usepackage{dcolumn}% Align table columns on decimal point
\usepackage{bm}% bold math
\usepackage{epsfig}
\usepackage{graphicx}% Include figure files
\usepackage{dcolumn}% Align table columns on decimal point
\usepackage{bm}% bold math

\newcommand{\mytilde}{\raise.17ex\hbox{$\scriptstyle\mathtt{\sim}$}}

\renewcommand{\bf}{\rm}

\title[Evidence for a Lower Value for $H_0$ from Cosmic Chronometers Data?]{Evidence for a Lower Value for $H_0$ from Cosmic Chronometers Data?} 

\author[V. C. Busti, C. Clarkson and M. Seikel]{Vinicius C. Busti$^{1}$\thanks{E-mail: 
vinicius.busti@uct.ac.za}, Chris Clarkson$^{1}$
and Marina Seikel$^{2,1}$
\\ $^{1}$  Astrophysics, Cosmology \& Gravity Centre (ACGC), and  \\ Department of Mathematics and Applied Mathematics, 
University of Cape Town, Rondebosch 7701, Cape Town, South Africa \\
$^{2}$ Physics Department, University of Western Cape, Cape Town 7535, South Africa} 

\begin{document}

\date{Accepted . Received ; in original form }

\pagerange{\pageref{firstpage}--\pageref{lastpage}} \pubyear{2013}

\maketitle

\label{firstpage}

\begin{abstract} 
\noindent 
An intriguing discrepancy emerging in the concordance model of cosmology is the tension between the locally measured value of the Hubble rate, and the `global' value inferred from the 
cosmic microwave background (CMB). This could be due to systematic uncertainties when measuring $H_0$ locally, or it could be that we live in a highly unlikely Hubble bubble, or other 
exotic 
scenarios. We point out that the global $H_0$ can be found by extrapolating $H(z)$ data points at high-$z$ down to $z=0$. By doing this in a Bayesian non-parametric way we can find a 
model-independent value for $H_0$. We apply this to 19 measurements based on differential age of passively
evolving galaxies as cosmic chronometers. Using Gaussian processes, we find $H_0=64.9 \pm 4.2$ km s$^{-1}$ Mpc$^{-1}$ $(1\sigma)$, in agreement with the CMB value, but reinforcing 
the tension with 
the local value. An analysis of possible sources of systematic errors shows that the stellar population synthesis model adopted may change the results significantly, being 
the main concern for subsequent studies. 
Forecasts for future data show that distant $H(z)$ measurements can be a robust method to determine $H_0$, where a focus in precision and a careful assessment of systematic errors
are required.
\end{abstract} 


%---------------------------------------------------------%
\section{Introduction}
%---------------------------------------------------------%

There is a strong tension, recently quantified by \cite{verde2013}, between the value of the Hubble constant $H_0$ derived by {\it Planck} \citep{planck} from anisotropies in the cosmic microwave 
background (CMB): $67.3 \pm 1.2$ km s$^{-1}$
Mpc$^{-1}$, and the value from local measurements: 
$73.8 \pm 2.4$ km s$^{-1}$ Mpc$^{-1}$ \citep{H0riess}. While the latter measurement is based on local measurements, the former 
infers a global value for the Hubble constant within a cosmological model. 

There remains disagreement about the local value of $H_0$ depending on the distance indicator used to measure it, which hints the discrepancy with {\it Planck} could be 
the result of systematic errors. \cite{H0riess} calibrated the SNe Ia distances with three indicators: distance to NGC 4258 based on a megamaser measurement, parallax
measurements to Milk Way cepheids (MWC) and cepheids observations and a revised distance to the Large Magellanic Cloud (LMC). 
Contrarily, calibrating the SNe Ia with the tip of red-giant branch, \cite{tammann} provides $H_0$ $=$ 63.7 $\pm$ 2.3 km s$^{-1}$ Mpc$^{-1}$. This shows how crucial is 
the first-step calibration in the distance ladder to measure $H_0$.

\begin{figure} 
%\vspace{.2in} 
\centerline{\epsfig{figure=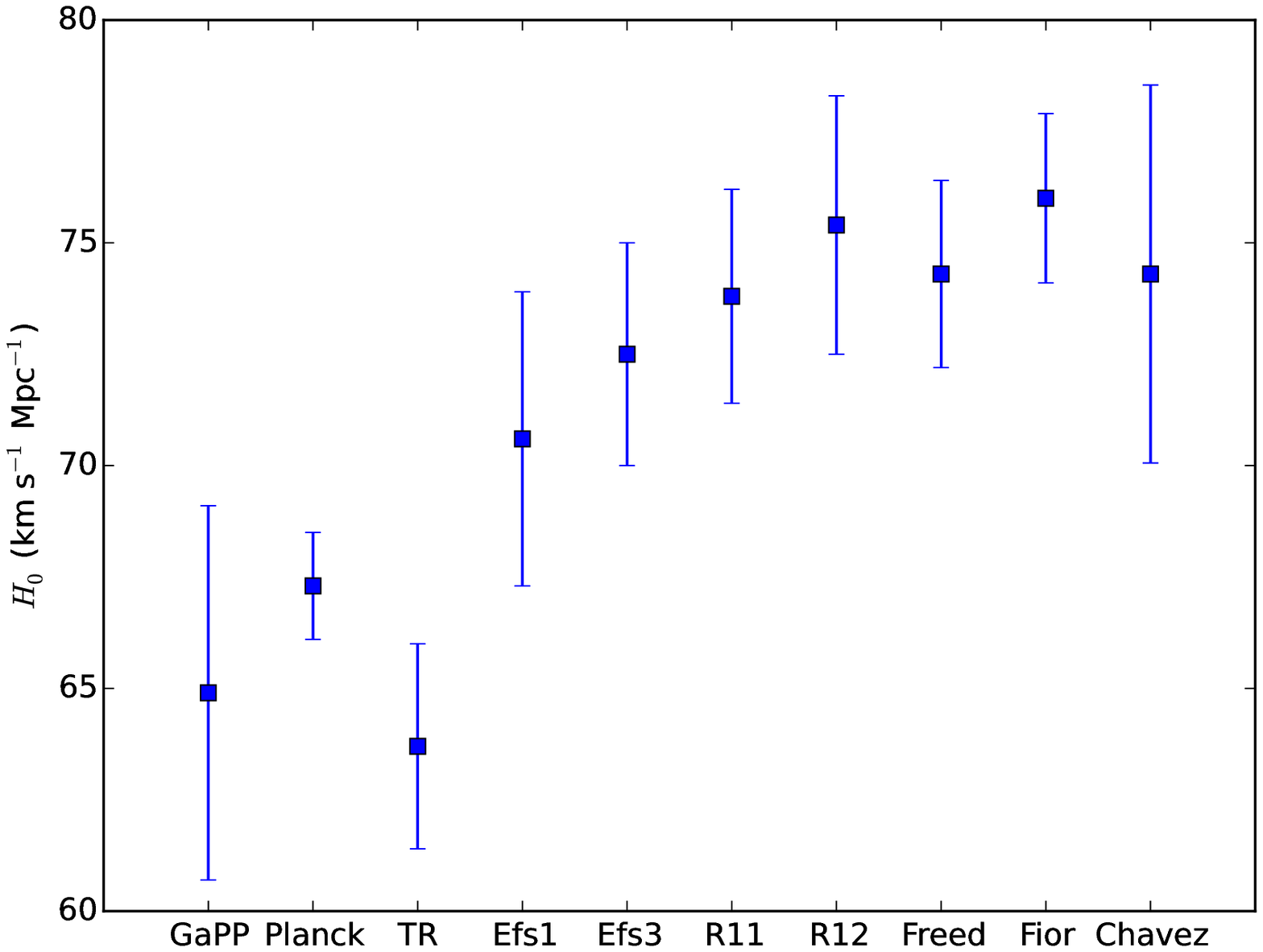,width=0.5\textwidth}
 \hskip 
0.1in} 
\caption{{\it Different measurements of $H_0$.} The figure shows how the result obtained in this work (GaPP) is compared to other determinations of $H_0$. The points refer to the 
following references: Planck \citep{planck}, TR \citep{tammann}, Efs1 \citep{efstathiou} with one anchor, Efs3 with three anchors, R11 \citep{H0riess}, R12 \citep{riess2012}, Freed \citep{freedman2012}, Fior \citep{fiorentino2013} and Chavez \citep{chavez2012}.} 
\label{figH0} 
\end{figure}

However, there are several 
local $H_0$ measurements with higher values. \cite{riess2012} found $H_0$ $=$ 75.4 $\pm$ 2.9 km s$^{-1}$ Mpc$^{-1}$ by using cepheids in M31. 
With a mid-infrared calibration for the cepheids, \cite{freedman2012} derived $H_0$ $=$ 74.3 $\pm$ 2.1 km s$^{-1}$ Mpc$^{-1}$, and with 8 new classical cepheids observed in galaxies
hosting SNe Ia \cite{fiorentino2013} got $H_0$ $=$ 76.0 $\pm$ 1.9 km s$^{-1}$ Mpc$^{-1}$.
By using HII regions and HII galaxies as distance indicators, \cite{chavez2012} obtained $H_0$ $=$ 74.3 $\pm$ 
3.1(random) $\pm$ 2.9 (syst.) km s$^{-1}$ Mpc$^{-1}$.  Some of these are over 4$\sigma$ away from the CMB-derived value. See Fig. \ref{figH0} for a plot of different 
measurements of $H_0$.

A variety of different physical effects could explain such a discrepancy. It could just be cosmic variance: as we can observe the Universe from only one position, we are not able to 
realize the global parameters from the local parameters, as in the local
expansion rate for instance. If we live in an locally underdense region, a ``Hubble bubble'', a higher value for $H_0$ is obtained compared to the global value. This effect was 
carefully addressed 
by \cite{cosmic_variance} through a modelling of the statistics of matter distribution which provides the distribution of the gravitational potential at the observer.
The outcome is that cosmic variance can alleviate the tension, but a complete elimination requires a very rare fluctuation \citep{cosmic_variance,cosmic_variance2}.

Another way to look at the problem is to consider that the discrepancy may indicate new physics, such as massive neutrinos \citep{hu}, or alternative dark energy models \citep{cde,xcdm}. 

Recently, some analyses were performed trying to identify sources of systematic errors in order to remove or alleviate the tension. For example, by using only the geometric maser 
distance to NGC 4258 of \cite{humphreys} as an anchor, \cite{efstathiou} revisited \cite{H0riess} analysis and derived $H_0=70.6 \pm 3.3$ km s$^{-1}$ Mpc$^{-1}$, while 
combining with LMC and MWC anchors the value is $72.5 \pm 2.5$ km s$^{-1}$ Mpc$^{-1}$, alleviating the tension. The {\it Planck} data were also reanalysed by \cite{spergel}, where it 
was claimed that the 217 GHz $\times$ 217 GHz detector is responsible for some part of the tension.
Their new Hubble constant without the 217 GHz $\times$ 217 GHz detector is slightly higher: $H_0 = 68.0 \pm 1.1$ km s$^{-1}$ Mpc$^{-1}$.

With so many alternatives, progress can be achieved by developing new ways to address the issue. We point out here that $H(z)$ data which are not calibrated on a $H_0$ estimate can be extrapolated to z=0 to provide 
an independent measurement of the global $H_0$. Here, the Hubble function is reconstructed in order to derive $H_0$ from 19 $H(z)$ 
measurements of passively evolving galaxies as cosmic chronometers \citep{jimloeb}. Many of these are at relatively moderate and high redshifts so intrinsically probe the global 
value for $H_0$ rather than the local one. We use Gaussian Processes (GP), which is a non-parametric method, to obtain the value of 
the Hubble constant in a completely cosmological model-independent way, which is in principle not affected by the local 
systematics. We show the value of the Hubble constant derived in this way is lower than the standard local measurements. We obtain $H_0$ $=$ $64.9 \pm 4.2$ km s$^{-1}$ Mpc$^{-1}$ $(1\sigma)$, in agreement with 
the CMB-inferred value. A better understanding of systematic errors, especially the adopted stellar population synthesis model, is required: we show that to improve this result a 
big effort is necessary to decrease the errors substantially in future, and a focus on precision is worthier than the 
number of data.

The paper is organized as follows: in Sec. \ref{meth} we describe GP as well as standard parametric methods adopted to constrain $H_0$. In Sec. \ref{const} the bounds 
derived for the Hubble constant are displayed, followed by forecasts of constraints in Sec. \ref{fore}. We finish the paper in Sec. \ref{conc}
with the conclusions.

%---------------------------------------------------------%
\section{Methods}
\label{meth}
%---------------------------------------------------------%

%---------------------------------------------------------%
\subsection{Gaussian Processes (GP)}
%---------------------------------------------------------%

A gaussian \emph{distribution} is
a distribution over random variables, while a gaussian \emph{process} is a distribution over functions. This allows one to reconstruct a function from data without assuming a parametrisation for it.  Here we use GaPP (Gaussian Processes in 
Python)\footnote{http://www.acgc.uct.ac.za/\mytilde seikel/GAPP/index.html} \citep{gapp} in order to reconstruct the Hubble parameter as a function of the redshift from which we can infer $H_0$.
This method has been applied for several purposes, for example the reconstruction of the equation of state of dark energy \citep{gapp} and to perform null tests of the concordance model
\citep{gapp1,gapp2}.

The reconstruction is given by a mean function with gaussian error bands, where the function values at different points $z$ and $\tilde{z}$ are connected through a  
covariance function $k(z,\tilde{z})$ (see \cite{sc2013} for a discussion of choices of covariance functions). This covariance function depends on a set of hyperparameters. 
Here, as we expect that the Hubble parameter and all its derivatives to
be smooth, we consider the general purpose squared 
exponential (Sq. Exp.) covariance function which is given by
\begin{equation}
 k(z,\tilde{z}) = \sigma_f^2 \exp\left\{-\frac{(z-\tilde{z})^2}{2l^2}\right\}.
\end{equation}
In the above equation we have two hyperparameters, the first $\sigma_f$ is related to typical changes in the function value while the second $l$ is related to the distance one needs
to move in input space before the function value changes significantly.
We follow the steps of \cite{gapp} and determine the maximum likelihood value for $\sigma_f$ and $l$ in order to obtain the value of the function. In this way, we are able to 
reconstruct the Hubble parameter as a function of the redshift from $H(z)$ measurements. 
We discuss in Sect. \ref{sect_cov} the impact of different covariance functions on our results.

%---------------------------------------------------------%
\subsection{Parametric Analyses}
%---------------------------------------------------------%

In order to compare the results provided by non-parametric methods with standard analyses, we also consider two parametric models. 
First of all, we take a flat XCDM model, where the universe is composed by dark matter and a fluid $X$ with equation of state $p_X=w \rho_X$, where the Hubble parameter is given by

\begin{equation}
  H(z)= H_0 \sqrt{\Omega_{\rm m}(1+z)^3 + (1-\Omega_{\rm m})(1+z)^{3(1+w)}},
 \label{xcdm}
 \end{equation}
where $\Omega_{\rm m}$ is the matter density parameter today. When $w=-1$ this is the concordance $\Lambda$CDM model which we consider separately.
In order to derive $H_0$ for the parametric models, we apply standard statistical procedures based on maximum likelihood methods.

%---------------------------------------------------------%
\section{Constraints on $H_0$}
\label{const}
%---------------------------------------------------------%

We use 19 $H(z)$ measurements \citep{hzmeasurements1,hzmeasurements2,hzmeasurements3} from passively evolving galaxies as cosmic chronometers to derive the value of $H_0$.  

\begin{figure} 
%\vspace{.2in} 
\centerline{\epsfig{figure=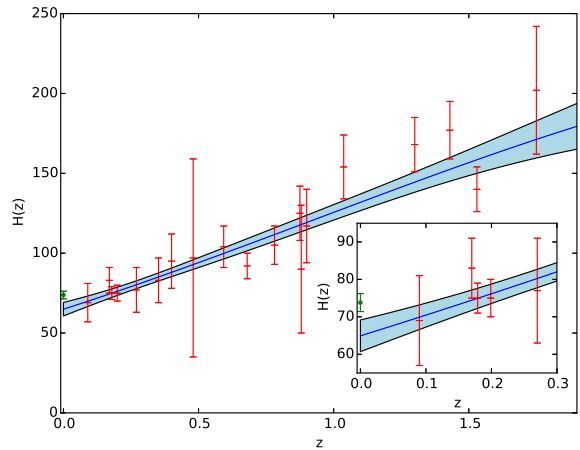,width=0.5\textwidth}
 \hskip 0.1in} 
\caption{Model independent reconstruction of $H(z)$ using Gaussian processes. The red points with error bars represent the 19 $H(z)$ measurements
 and the blue shaded contour the reconstruction within the $1\sigma$ confidence level. 
 For comparison purposes, we also show the value obtained by Riess et al. (2011), where we see that it is compatible 
to the GaPP value only at 2$\sigma$. The inset shows a zoom in the low redshift region.} 
\label{fig1} 
\end{figure}

Figure \ref{fig1} presents the results for the non-parametric approach adopted in this work. We also plot the $H(z)$ measurements with their respective
errorbars. The blue solid line refers to the reconstruction with GaPP, with
the shaded contours designating the $1\sigma$ errors. When extrapolated to  redshift $z=0$, we obtain 
$H_0 = 64.9 \pm 4.2$ km s$^{-1}$ Mpc$^{-1}$ $(1\sigma)$. Note that this value is completely independent of a cosmological model, which makes it complementary but consistent with 
the {\it Planck} value which is derived within the $\Lambda$CDM model. In this way, our result can also be used to shed light in the whole cosmological model. This result goes in the 
direction of a model-independent approach with SNe Ia which also prefers lower 
values for $H_0$ \citep{emille}.

As a means to compare the non-parametric with standard parametric analyses, a flat $\Lambda$CDM model and a flat XCDM are also considered. The Hubble
constant is found  to be $H_0 = 68.9 \pm 2.8$ km s$^{-1}$ Mpc$^{-1}$ $(1\sigma)$ for the flat $\Lambda$CDM model and 
$H_0 = 69.0 \pm 6.7$ km s$^{-1}$ Mpc$^{-1}$ $(1\sigma)$ for XCDM. The bigger error for the XCDM case is derived as a consequence of the inclusion of an extra parameter.
The mean values are higher compared to the non-parametric approach, and in closer agreement with locally measured values, but also in agreement with the non-parametric result.

Table \ref{table1} summarizes the constraints for $H_0$ with $1\sigma$ errors. All methods prefer values for $H_0$ below 
70 km s$^{-1}$ Mpc$^{-1}$, in  contrast with some local determinations of $H_0$.

\begin{table}
\caption{$H_0$ constraints from 19 $H(z)$ measurements.}
\label{table1}
\begin{center}
\begin{tabular}{@{}cccc@{}}
\hline Method & $H_0$ $\pm$  $1\sigma$ & $\sigma_{H_0}$   \\
          & (km s$^{-1}$ Mpc$^{-1}$)   &  (km s$^{-1}$ Mpc$^{-1}$)   
\\ \hline\hline

Sq. Exp.                    &  64.9  &    4.2     \\
Mat\'{e}rn$(9/2)$           &  65.9  &    4.5     \\
Mat\'{e}rn$(7/2)$           &  66.4  &    4.7     \\
Mat\'{e}rn$(5/2)$           &  67.4  &    5.2     \\
$\Lambda$CDM                &  68.9  &    2.8     \\
XCDM                        &  69.0  &    6.7     \\

\hline
\end{tabular}
\end{center}
\end{table}

%-------------------------------------------------------------------%
\subsection{Systematic Errors}
%-------------------------------------------------------------------%

Some tests were performed in order to evaluate the robustness of the result. We split our analysis searching for three effects: (i) the impact of the covariance function 
in GaPP, (ii) a possible presence of outliers driving $H_0$ for lower values and (iii) systematic errors from the stellar population synthesis (SPS) models.

%-------------------------------------------------------------------%
\subsubsection{Covariance Functions}
\label{sect_cov}
%-------------------------------------------------------------------%

The freedom in the GP approach comes in the covariance function. While in traditional parametric analyses we choose a model to characterise what is our prior belief
about the function in which we are interested, with GP we ascribe in the covariance function our priors about the expected function properties (e.g. smoothness, correlation scales etc.).

Since we expect the Hubble parameter and its derivatives to be smooth, the squared exponential covariance function was selected which is infinitely differentiable~-- this implies 
that functions drawn from the process are also infinitely differentiable. 
However, we considered other covariance functions to see how the results are affected. In order to do so, we considered three covariance functions 
from the Mat\'{e}rn family, namely the $\nu=5/2$, $7/2$ and $9/2$ (see \cite{sc2013} for definitions and further discussion). Writing $\nu = p + 1/2$, each Mat\'{e}rn function is $p$ 
times differentiable as are functions drawn from it, and
the squared exponential is recovered for $\nu \rightarrow \infty$. Increasing $\nu$ increases the width of the covariance function near the peak implying stronger correlations from 
nearby points for a fixed correlation length $\ell$.

The results are shown in Table \ref{table1}, where we see slightly higher values are derived for $H_0$, together with slightly larger errors, for smaller $\nu$. In fact, for the 
Mat\'{e}rn$(5/2)$ the tension with local $H_0$ disappears, although the result remains 
in fully agreement with {\it Planck}. Interestingly, although there is some shift, the errors are relatively independent of the covariance function choice, especially when compared to 
the ones derived when one increases the number of parameters in parametric analyses (the error more than doubles when allowing for $w$ to be free, compared to fixing it to $-1$), 
showing that GP provide very stable results within different reasonable assumptions.

%-------------------------------------------------------------------%
\subsubsection{Presence of Outliers}
%-------------------------------------------------------------------%

We checked if the high-redshift data were pivoting down the value of $H_0$ to smaller values. To do so, we removed 
all data points with redshifts greater than 1, but again the results were completely consistent with the full sample, 
$H_0 = 66.9 \pm 4.3$ km s$^{-1}$ Mpc$^{-1}$, thus not removing the tension. By removing low-redshift points, first and second
or third and fourth, again the results did not change significantly, with $H_0 = 66.3 \pm 4.6$ km s$^{-1}$ Mpc$^{-1}$ and $H_0 = 66.4 \pm 7.1$ km s$^{-1}$ Mpc$^{-1}$, respectively.  
We note that the points with smaller errors dominate the final error budget, as confirmed also by the analysis done in Sect. 4.

We also removed point by point in the analysis. For the first 17 points, the results changed slightly, with a mean value between 64 and 65 and errors between 4 and 5. Conversely, 
the high-redshift points showed the biggest departure once removed, with values 
$H_0 = 71.5 \pm 5.9$ (18th out) and $H_0 = 73.2 \pm 8.7$ km s$^{-1}$ Mpc$^{-1}$ (19th out). Higher values are derived and the error blows up, although in agreement with the full 
sample. This shows how $H_0$ is sensitive to high-redshift values, where more data points in this redshift region might help to mitigate possible sytematic errors due to outliers.   

%-------------------------------------------------------------------%
\subsubsection{Different SPS Models}
%-------------------------------------------------------------------%

One of the possible main sources of systematic errors in $H(z)$ measurements comes from the adopted SPS model. The 19 points used here were derived with \cite{bc03} SPS model (BC03). 
On the other hand, recently \cite{hzmeasurements3} calculated $H(z)$ for eight measurements considering BC03 and 
another SPS model from \citet[MaStro]{mastro}.
We performed the 
reconstruction with GaPP for this subset with both SPS models. For BC03 we derived $H_0$ in the range $64.4 \pm 4.9$ km s$^{-1}$ Mpc$^{-1}$, in good agreement with the full sample. 
On the contrary, the analysis with MaStro provided $H_0 = 75.1 \pm 5.2$ km s$^{-1}$ Mpc$^{-1}$, in disagreement with {\it Planck} and in good agreement with the value of \cite{H0riess}. 
Therefore, even with only eight data points, we identify the SPS model as the main concern for our results.

%-------------------------------------------------------------------%
\subsection{Other Data Sets}
%-------------------------------------------------------------------%

Another independent measurement for $H(z)$ is given by baryon acoustic oscillations (BAOs). Currently, there are 7 measurements from \cite{blake}, 
\cite{reid}, \cite{xu}, \cite{busca} and \cite{wang}. Combining with the other measurements, for GaPP (Sq. Exp.) we got $H_0 = 69.4 \pm 4.4$ km s$^{-1}$ Mpc$^{-1}$,
for a flat $\Lambda$CDM  model $H_0 = 68.4 \pm 2.0$ km s$^{-1}$ Mpc$^{-1}$ and for a flat XCDM model $H_0= 69.8 \pm 4.6 $, all values consistent with {\it Planck}. 
However, there are some drawbacks when using the BAO data. 
First, these data are not model independent. They are based on the $\Lambda$CDM model to study the correlation functions and transform them to the $H(z)$ values. Moreover, generally
what is inferred is the combination $Hr_s$, $r_s$ standing for the sound horizon whose value is given by {\it WMAP} 
\citep{wmap5,wmap7,wmap9}.
Also important is the point raised by \cite{blake} warning that their values of $H(z)$ are derived and so they should not be used to test models.

Since the current errors do not allow a final decision about 
the value of $H_0$, our next step is to study whether future data can settle the issue.

%---------------------------------------------------------%
\section{Forecasts}
\label{fore}
%---------------------------------------------------------%

The procedure to analyse how future data can improve the determination of $H_0$ is split in two ways. First, we consider how the increase
of data points of same quality can change the constraints. Second, the errors for $H(z)$ are shrunk and a comparison is made between the number
of data and their quality.

The current errors for $H(z)$ measurements grow with redshift a few percent up to around 15 per cent. Assuming that future data will provide measurements with the same errors, 
we update the method of \cite{ma} to predict future data based on the recent measurements from \cite{hzmeasurements3}.
A value for $H(z)$ is generated by $H_{sim}(z) = H_{fid}(z) + {\cal{N}}(0,\tilde{\sigma}(z))$,
where $H_{sim}(z)$ and $H_{fid}(z)$ are respectively the simulated and fiducial values for the Hubble parameter at redshift $z$, and
${\cal{N}}(0,\tilde{\sigma}(z))$ is a random number gaussianly distributed with mean zero and variance $\tilde{\sigma}(z)$. To estimate 
$\tilde{\sigma}(z)$, the uncertainties of the observational points
are restricted by two straight lines: $\sigma_{+}=15.76z+3.65$ and $\sigma_{-}=13.29z+1.62$, with two ``outliers'' removed since they were not following the trend of the errors. 
Assuming that the errors of future data will be between the two lines,
one can expect the mean line of the error to be $\sigma_{0}=14.52z+2.63$. Therefore, the error of the simulated point is drawn from a gaussianly
distributed random variable $\tilde{\sigma}(z)= {\cal{N}}(\sigma_{0}(z),\epsilon(z))$, where $\epsilon(z) = (\sigma_{+}-\sigma_{-})/4$ is chosen 
to assure the error is within $\sigma_{-}$ and $\sigma_{+}$ with 95.4\% probability.

\begin{figure} 
%\vspace{.2in} 
\centerline{\epsfig{figure=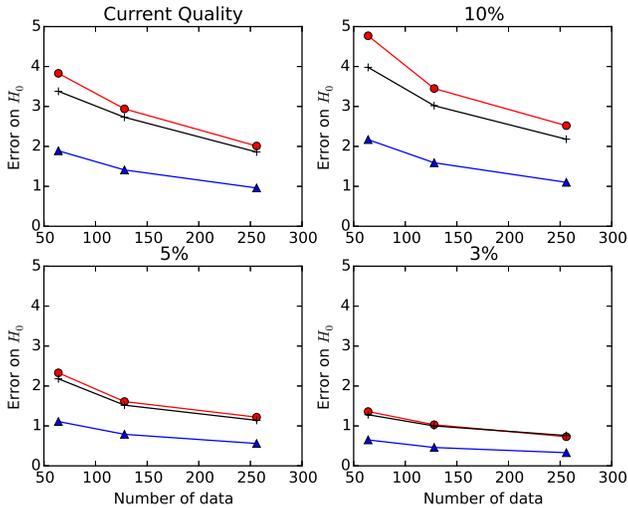,width=0.55\textwidth}
 \hskip 
0.1in} 
\caption{\emph{Forecasts for errors on $H_0$ (in km s$^{-1}$ Mpc$^{-1}$) with different number of data points.} The future data is evenly spaced in the redshift range 
$0.1 \leq z \leq 1.8$. The black crosses 
refer to the error provided by GaPP, while the red points refer to a flat XCDM model and the blue triangles to a flat $\Lambda$CDM model. The upper left panel refers to simulated
data with the same quality as current data. The upper right panel for simulated data with 10\% precision, the lower left with 5\% and the lower right with 3\% precision.}

\label{fig2} 
\end{figure}

Figure \ref{fig2} presents the expected future errors from considering data are equally spaced in the interval $0.1 \leq z \leq 1.8$.  
We present the expected error on $H_0$ from simulations with 64, 128, and
256 data points. These numbers were chosen because with 64 $H(z)$ measurements of same quality as today one can achieve the same constraints given by current SNe Ia \citep{ma}. 
The black crosses represent the errors with the GaPP reconstruction, the red dots for the flat XCDM model and the blue triangles for
the flat $\Lambda$CDM model. The first panel in the left shows the behavior of future data with the same quality as today, and 
the others show the trend for smaller errors for the $H(z)$ data, of 10\%, 5\%, and 3\% (see \cite{crawford} for an observational program to achieve such values).

Some conclusions can be made from Fig.~\ref{fig2}:

\begin{itemize}
 \item For future data with the same quality as today GaPP performs very well, with errors smaller than the ones obtained with a flat XCDM model. 
 
 \item Current quality data provide better constraints to $H_0$ than a constant error of 10\% in the whole redshift range, showing that lower redshift objects with higher 
precision compensate the low quality data at high redshifts.  
 
 \item For higher precision measurements GaPP and the XCDM model provide the same constraints, showing that a non-parametric approach is powerful to study cosmological data. 

 \item Improvement in precision is more important than increasing the number of data of poorer quality.
\end{itemize}

%---------------------------------------------------------%
\section{Conclusions}
\label{conc}
%---------------------------------------------------------%

We have applied GaPP, a non-parametric smoothing method based on Gaussian Processes, to 19 $H(z)$ measurements in order to constrain the Hubble constant $H_0$.
This method does not rely on a cosmological model, so its results can be used to infer the impact of systematic errors as well as the underlying cosmological framework. 
We have obtained $H_0$ to be $64.9 \pm 4.2$ km s$^{-1}$ Mpc$^{-1}$ $(1\sigma)$, a value which is in agreement with {\it Planck}, but in disagreement with local measurements. This supports 
the notion that either there are unidentified systematic errors in the local $H_0$ data, or the local value is indeed different from the global value. 
A better comprehension of systematic errors, especially a thorough analysis of the impact of SPS models, can improve the robustness of our results.
Simulations have shown that improvements in distant $H(z)$ measurements can help pin down the global value of $H_0$.

%----------------------------------------------------------%
\section*{Acknowledgments}
%----------------------------------------------------------%

The authors are grateful to Raul Jimenez and Martin Hendry for useful comments.
VCB is supported by CNPq-Brazil through a fellowship within the program Science without Borders. MS is supported by the South Africa Square Kilometre 
Array Project and the South African National
Research Foundation (NRF).

\label{lastpage}

\end{document}